\documentclass[10pt,letterpaper]{article}
\usepackage{opex3}
\usepackage{amssymb}

\begin{document}
\title{Efficient generation of $>$2 W of green light by single pass frequency doubling in PPMgLN}

\author{M. G. Pullen, J. J. Chapman and D. Kielpinski}

\address{Center for Quantum Dynamics, Griffith University, \\ Brisbane, Australia}

\email{mgpullen@gmail.com}


\begin{abstract*} 

We report 32\% efficient frequency doubling of single frequency 1029 nm light to green light at 514.5 nm using a single pass configuration.  A congruent composition, periodically poled magnesium doped lithium niobate (PPMgLN) crystal of 50 mm length was used to generate a second harmonic power of 2.3 W.  To our knowledge, this is the highest reported frequency doubling efficiency of any wavelength light in a PPMgLN crystal and also the highest reported SHG output power in the green for PPMgLN.\end{abstract*}



\section{Introduction}
Second harmonic generation (SHG) is important for producing visible light at wavelengths for which lasers are not available. Two different configurations have been used to achieve frequency doubling: intracavity \cite{Hayasaka04} and single pass \cite{Pavel04} systems.  While intracavity designs tend to achieve higher efficiencies than single pass setups, they are significantly harder to construct and take a longer time to implement.  Highly efficient single pass SHG is therefore desirable for many industrial applications.  Typical crystals used in frequency doubling experiments include potassium titanyl phosphate (KTP), lithium triborate (LBO) and lithium niobate (LN).  Achieving good phase matching conditions in these materials can be quite challenging as they have very strict index matching requirements. To overcome this challenge, quasi phase matching (QPM) \cite{Somekh72} was suggested by creating periodically poled crystals using the ferroelectric properties of both KTP and LN.  These crystals are denoted PPKTP and PPLN respectively and increase the achievable SHG efficiency.  Periodically poled crystals still have limitations however, as at medium powers PPLN crystals can suffer from pointing instability and at high powers permanent photorefractive damage can become an issue \cite{Miller97}.  To reduce these effects, periodically poled lithium niobate crystals are often doped with magnesium oxide (PPMgLN) \cite{Bryan84}.  The first demonstration of PPMgLN being used for frequency doubling was the generation of 1 mW of 437 nm blue light \cite{Mizuuchi96}. Since then, low power SHG of light at wavelengths as low as 373 nm \cite{Mizuuchi03} and 340 nm \cite{Mizuuchi032} have also been reported in PPMgLN.

Previously, reports of green light generation by frequency doubling have mostly been made at the 532 nm \cite{Miller97} and 540 nm \cite{Ou92} wavelengths.  Very limited reports \cite{Guo06} have been made of the SHG of light in the 515 nm region. High power light in this region can be useful in display technologies, as a substitute for argon ion lasers and as pump sources for tunable lasers. In this paper we report the generation of 2.3 W of 514.5 nm green light via SHG in a single pass configuration. A maximum efficiency of 32\% was achieved and to our knowledge, this is the highest reported frequency doubling efficiency of any wavelength light in a PPMgLN crystal and also the highest reported SHG output power in the green for PPMgLN.\\

\section{Experiment}

\begin{figure}[!htp]
	\centering
		\includegraphics[width=125mm]{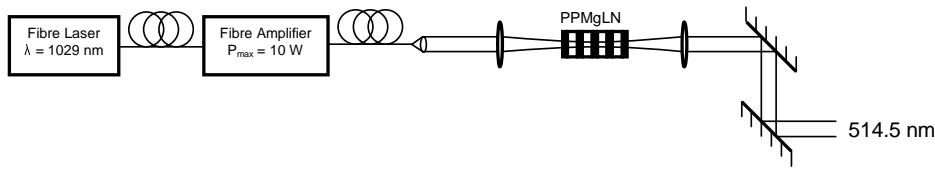}
			\caption{The experimental configuration used in our measurements.  A single frequency fibre seed laser is amplified in a ytterbium doped fibre amplifier and focused into a PPMgLN crystal.  The 1029 nm and 514.5 nm wavelengths are separated via two dichroic mirrors and the green output is monitored with a power meter.}
	\label{fig:PullenFig1}
\end{figure}

Our experimental setup is illustrated in Fig. \ref{fig:PullenFig1}.  The frequency doubling crystal (HC Photonics) is PPMgLN with a MgO doping concentration of 5\% and is of congruent composition. PPMgLN was chosen as the frequency doubling crystal as it minimises both photorefractive damage \cite{Pavel04} and green induced infrared absorption (GRIIRA) \cite{Furukawa01}. The dimensions of the crystal are 0.5 $\times$ 3 $\times$ 50 mm and the QPM period is 6.15 $\mu\mathrm{m}$ with a 50\% duty cycle.  The input and output surfaces have been anti-reflection (AR) coated for 1029.3 nm and 514.65 nm respectively.

\begin{figure}[!htp]
	\centering
		\includegraphics[width=125mm]{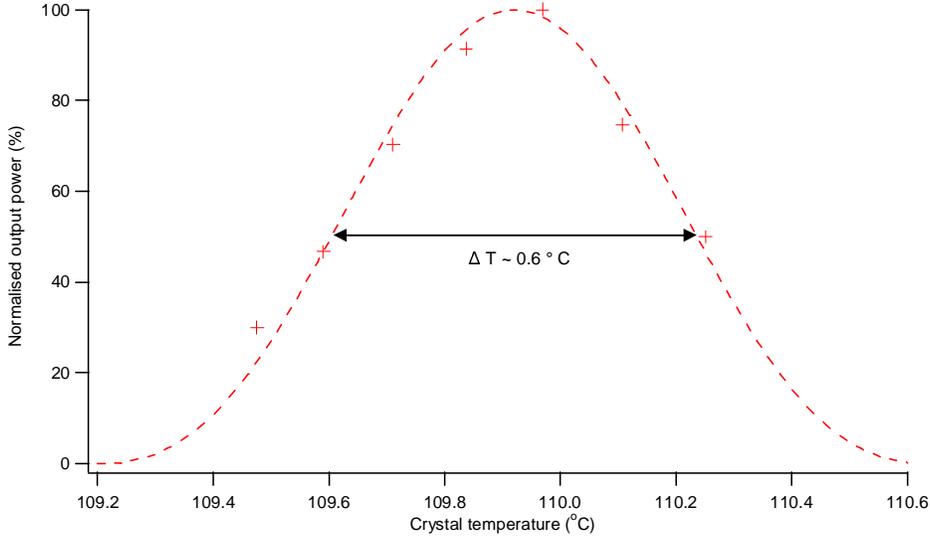}
			\caption{Oven temperature versus output intensity showing a FWHM phase matching temperature bandwidth of $\sim\mathrm{0.6}\,^{\circ}\mathrm{C}$.  The crosses show actual data points while the dashed line is a fit to a sinc$^2$ function.}
	\label{fig:PullenFig2}
\end{figure}

The PPMgLN crystal is housed in an externally controlled, temperature stabilised oven which was supplied by the crystal manufacturer.  The oven has $0.1\,^{\circ}\mathrm{C}$ precision and during normal operation provides the required phase matching temperature of $109.9\,^{\circ}\mathrm{C}$ for sustained periods of time ($> \mathrm{48}$ hours).  Due to the large length of the crystal, the FWHM phase matching temperature bandwidth is $\sim\mathrm{0.6}\,^{\circ}\mathrm{C}$ as shown in Fig. \ref{fig:PullenFig2}.

The seed laser is a single frequency linearly polarised fibre laser which has an output power of 6 mW at 1029 nm.  The output of this laser propagates through single mode polarisation maintaining fibre and is used as the input to an IPG Photonics fibre amplifier (YAR-10K-1030-LP-SF).  The amplifier provides a linearly polarised output with a maximum power of 10 W at 1030 nm.  A lens is then used to focus the beam waist into the centre of the crystal. The generated green light is separated from the IR light after the crystal via two dichroic mirrors with high reflectivity in the green and high transmission in the IR. The IR light is incident on a beam stop while the green light power is monitored by a Coherent FieldMate power meter.  We used the Fresnel reflection from a thick glass plate to pick off 1\% of the output green light. This beam was allowed to diverge over approximately 2 m, forming a large image of the beam, which was used for continuous monitoring of the beam profile during our experiments.

\section{Results}

Our initial attempts at SHG generation were performed with low input powers ($\sim\mathrm{1 W}$) to assess crystal performance and to limit the possibility of damage.  During these initial tests, it was observed that phase matching conditions were only being maintained for a matter of minutes.  As the oven only heated the underside of the crystal, a temperature differential higher than the phase matching bandwidth was created over it's width which could cause frequency doubling instabilities.  After the construction of a foam enclosure for the oven which provided a more consistent temperature, phase matching conditions could be maintained for much longer periods of time with a stable 190 mW output at a normalised conversion efficiency of 13.2 \%/W.  We believe that better temperature stability could be achieved by heating the crystal on both the top and bottom faces.

\begin{figure}[!htp]
	\centering
		\includegraphics[width=125mm]{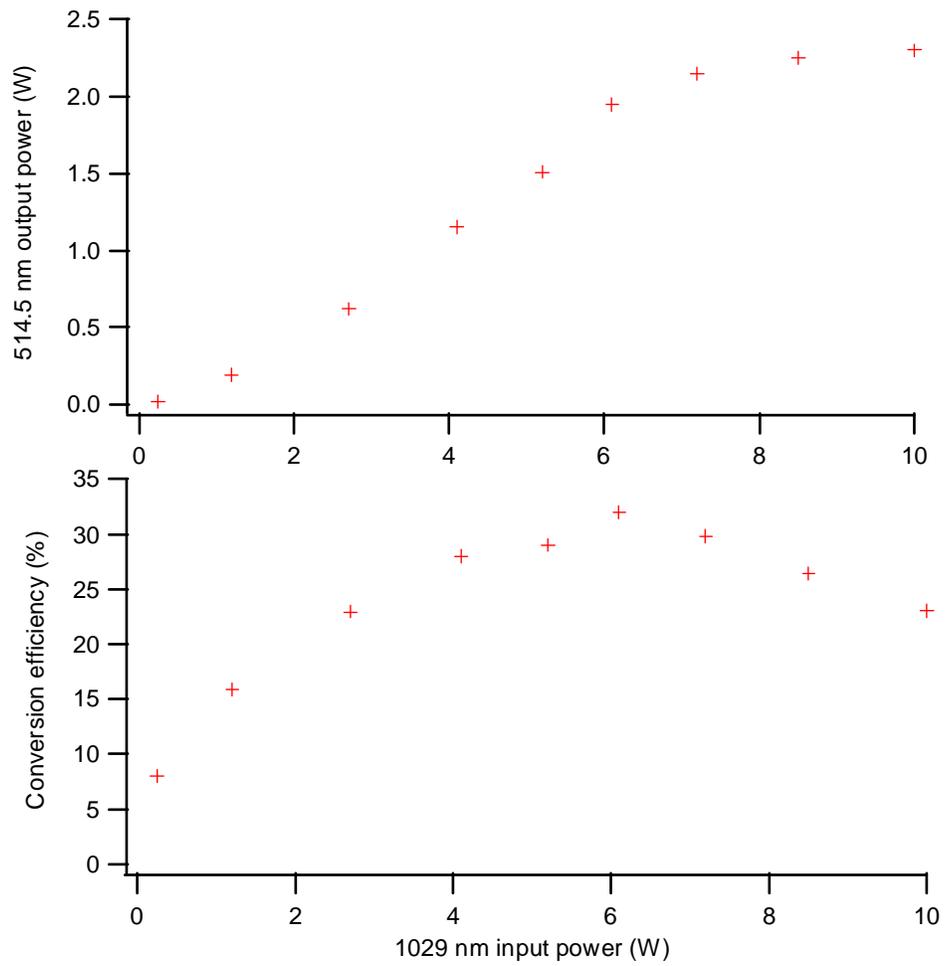}
			\caption{Top panel: Green output power versus IR input power at an IR beam waist of $\sim$55 $\mu\mathrm{m}$. Bottom panel: SHG conversion efficiency. We do not correct for residual reflections from the crystal end facets.}
	\label{fig:PullenFig3}
\end{figure}

Fig. \ref{fig:PullenFig3} shows the SHG output power versus the input power at an IR beam waist of $\sim$55 $\mu\mathrm{m}$, the maximum achieved green power was 2.3 W at a conversion efficiency of 23\%.  It is necessary to stress that all quoted measurements are actual measured powers, uncorrected for residual reflection at the crystal facets.  The results show that the output power starts to reach a plateau near 2 W, the conversion efficiency at this stage reaches a maximum of 32\% and starts to decrease. GRIIRA can be expected to have a significant effect on SHG for $\alpha^{-1} \lesssim L_{C}$, where $L_{C}$ is the crystal length and $\alpha$ is the crystal IR absorption coefficient.  Rough calculations using previously reported GRIIRA data for PPMgLN \cite{Furukawa01} and derived SHG relations \cite{Boyd68} yield an estimated absorption coefficient between 0.1 cm$^{-1}$ and 0.3 cm$^{-1}$ at the maximum green power.  In this case our 50 mm crystal should suffer from significant IR absorption.  We therefore believe that the saturation effect and hence decrease in conversion efficiency seen in Fig. \ref{fig:PullenFig3} is a direct result of GRIIRA.  

It was observed that as the input power was increased above 7 W, the stability of the green output power level declined and the beam profile shifted from its normal Gaussian shape.  At these higher input powers, we observed what seemed to be horizontal contours of constant intensity.  As the crystal was mounted with no thermal contact on the upper side, most thermal gradients would be expected to be in that direction.  We therefore attribute this behaviour to phase-matching problems due to thermal instability.  In addition to these horizontal temperature gradients, GRIIRA may also be able to create local thermal gradients which are caused by varying amounts of IR absorption. When the input powers were again lowered below 7 W, stability could be easily achieved which suggests that no permanent crystal damage had occurred in contrast to reports for undoped PPLN \cite{Miller97}.  It is also important to note that during high input power measurements we detected no significant change of the overall beam size, suggesting that no photorefraction was present.

\section{Summary}

In summary, we report 32\% efficient SHG of single mode green light at 514.5 nm using a single pass configuration.  PPMgLN crystal of congruent composition was used to generate a maximum second harmonic power of 2.3 W which we believe was limited by GRIIRA and thermal gradients.  To our knowledge, this is the highest reported frequency doubling efficiency of any wavelength light in a PPMgLN crystal and is also the highest reported SHG output power in the green for PPMgLN.
\\
\\
\noindent
\textbf{Acknowlegments}
\\
\\
\noindent
This work was supported by the US Air Force Office of Scientific Research under contract FA4869-06-1-0045 and Prof. Howard Wiseman's Australian Research Council Federation Fellowship grant FF0458313.

\end{document}